\documentclass[a4paper,11pt]{article}         
\usepackage{amsmath,amssymb,amsfonts}    
\usepackage{graphicx,caption}            
\usepackage{color}                       
\usepackage{hyperref}                    

\usepackage{makeidx}                  
\usepackage{acronym}
\usepackage{mathrsfs}
\usepackage{float}
\usepackage{vruler}                     
\usepackage{palatino, url, multicol}
\usepackage{setspace}                    
\usepackage{txfonts}

\usepackage{tikz}
\usetikzlibrary{shapes.geometric, arrows}
\usetikzlibrary{calc}

\usepackage[numbers,sort&compress]{natbib}   
\usepackage{import}

\usepackage{caption}

\usepackage{listings}

\usepackage{comment}

\restylefloat{figure}

\begin{document}

\title{Void Shape Identification in a 2D Point Distribution}
\author{\Large{Netzer Moriya}}

\date{}

\maketitle

\section{Abstract}

We introduce a new approach for identifying and characterizing voids within two-dimensional (2D) point distributions through the 
integration of Delaunay triangulation and Voronoi diagrams, combined with a Minimal Distance Scoring algorithm.

Our methodology initiates with the computational determination of the Convex Hull vertices within the point cloud, followed 
by a systematic selection of optimal line segments, strategically chosen for their likelihood of intersecting internal void 
regions. We then utilize Delaunay triangulation in conjunction with Voronoi diagrams to ascertain the initial points 
for the construction of the maximal internal curve envelope by adopting a pseudo-recursive approach for higher-order void identification. 
In each iteration, the existing collection of maximal internal curve envelope  points serves as a basis for identifying additional 
candidate points. 
This iterative process is inherently self-converging, ensuring progressive refinement of the void's shape with each successive 
computation cycle. 
The mathematical robustness of this method allows for an efficient convergence to a stable solution, reflecting both the 
geometric intricacies and the topological characteristics of the voids within the point cloud.

We apply our method to two distinct point set ensembles: a canonical single circle with random dispersion along the circumference, and a 
pair of congruent circles with an intentionally created internal void. In both examined cases, the method effectively unveils the internal 
void shape, pinpointing key vertices that delineate its contours. Additionally, this approach allows for a degree of control over the 
intricacies of the internal shape, demonstrating its refining in capturing geometric details.

Our findings introduce a method that aims to balance geometric accuracy with computational practicality. The approach is designed to 
improve the understanding of void shapes within point clouds and suggests a potential framework for exploring more complex, 
multi-dimensional data analysis.

\section{Introduction}

Given a finite set $P = \{p_1, p_2, \ldots, p_n\}$ where each point $p_i \in P$ is a tuple $(x_i, y_i)$ representing spatial coordinates in a 
two-dimensional Euclidean space, often characterized by irregular distributions and high-dimensional variability ("cloud"), we aim to 
identify and analyze the shape of a void within the cloud.

The identification of voids, cavities, or regions of sparse point density within a point cloud is a pivotal inverse geometry problem with 
significant implications in computational geometry \cite{bubenik:2012,Chalmoviansky:2003}, computer vision \cite{Jiang:2022}, 
computer graphics \cite{Salvaggio:2013,deng:2021} and robotics \cite{Karim:2000}. 
This multifaceted challenge requires an intricate balance between geometric precision and computational practicality \cite{Salvaggio:2014}. 

Recently, the importance of void shape analysis in point clouds is increasingly recognized, especially in areas like 3D modeling 
and reconstruction \cite{Fei:2022, Hung:2006} underlines the growing interest in not just detecting but also understanding the shape and 
characteristics of voids within point clouds \cite{Hoyle:2002}.

Traditional methods for void identification in point clouds, primarily density-based techniques \cite{Hu:2023}, focus on calculating point 
densities to pinpoint potential voids in lower density areas \cite{Srivastava:2009}. These methods, including analysis-by-synthesis 
approaches and nearest neighbor calculations \cite{Nguyen:2015}, often face challenges with varying density distributions and 
distinguishing between noise and sparse regions. Graph theory and clustering algorithms like DBSCAN \cite{Schubert:2017} provide 
insights into point connectivity but may not capture complex spatial relationships inherent in point clouds \cite{Karim:1998}. 
Additional geometrical approaches, such as the Ball Pivoting mesh generation algorithm, offer alternative methods for void 
identification \cite{Wright:2017, Bird:2019}, yet they still exhibit limitations in characterizing the intricate details of 
void structures in refined settings.

To bridge this gap, our approach leverages the mathematical intricacies of Delaunay triangulation~\cite{Mavriplis:1995} alongside the 
spatial analysis capabilities of Voronoi diagrams \cite{Okabe:2008}. This integration is further enhanced by a new method designed to 
reconstruct the internal shape of voids using a pseudo-recursive technique. Delaunay triangulation is renowned for its distinctive property,
where no point in a set lies within the circumcircle of any triangle in the triangulation. This offers a robust framework for delineating 
interstitial relationships within a point cloud. Concurrently, the Voronoi diagram, acting  as a geometric complement to Delaunay 
triangulation, counterpart to Delaunay triangulation, provides a natural and intuitive means for assessing point density and spatial 
interconnections.

Integrating these classical geometric constructs with a robust algorithm inspired by convex hull algorithms \cite{Preparata1977ConvexHulls}, 
our methodology aims to provide a nuanced understanding of voids within point clouds. This approach reconciles the need for geometric 
accuracy with the inherent variability and complexity of point cloud data, facilitating advanced spatial analysis. We acknowledge the 
potential computational challenges, such as time complexity and scalability, and aim to address these while balancing precision and 
practicality.

\section{Problem Description: Maximal Internal Envelope (MIE)}

Consider a finite set $P = \{p_1, p_2, \ldots, p_n\}$, where each $p_i \in \mathbb{R}^2$ represents a point in two-dimensional Euclidean 
space, denoted as $p_i = (x_i, y_i)$. Our goal is to identify an optimal curve $C$ that accurately delineates the boundary of a low-density 
region or void within the point cloud. The void is conceptualized as a region in $\mathbb{R}^2$ characterized by a low point density.

Given:
\begin{itemize}
    \item A point cloud $P$ in a two-dimensional space.
    \item A local density function $\rho$, derived from Delaunay triangulation and Voronoi diagrams. Delaunay triangulation is used to 
	establish connectivity between points, and the corresponding Voronoi diagram provides insights into the local density and distribution 
	of points. Specifically, larger Voronoi cells indicate lower local densities, which are utilized to define the function $\rho$.
	
\end{itemize}

Objective:
\begin{itemize}
    \item To identify an optimal curve $C$ that encapsulates the void. The curve should be closed, non-self-intersecting, and optimized 
	to maximize the enclosed area while minimizing its length. The curve's shape, not necessarily convex, reflects the complex nature of 
	void structures in point clouds.
\end{itemize}

Constraints:
\begin{equation}
    C = \underset{C'}{\mathrm{argmin}} \left\{ \mathrm{Length}(C') \; \Bigg| \; 
    \begin{aligned}
        & C' = \bigcup_{i=1}^{n} [P_i, P_{i+1}], P_{n+1} = P_1, \text{ satisfying:} \\
        & \quad \begin{array}{@{}l@{}}
            1. \, [P_i, P_{i+1}] \text{ are linear segments, } \forall i \in \{1, \ldots, n\} \\
            2. \, \forall i \neq j, \, [P_i, P_{i+1}] \cap [P_j, P_{j+1}] = \emptyset \\
            3. \, \text{Void is enclosed by } C', \text{ no segments intersect the void} \\
            4. \, \max \left\{ \text{Area enclosed by } C' \right\} \text{ for a given } \mathrm{Length}(C')
          \end{array}
    \end{aligned}
    \right\}
\end{equation}

subject to:
\begin{align}
    C' &\text{ maximizes the fidelity to the void's shape}, \\
    C' &\text{ minimizes the deviation from the points defining the boundary of the void}, \\
    C' &\text{ balances the trade-off between simplicity and accuracy}.
\end{align}

This problem encapsulates the geometric and computational complexities of identifying and accurately delineating low-density regions 
within point clouds.

\section{Proposed Methodology}

The proposed methodology for elucidating the intricate shape characteristics of voids within a two-dimensional point cloud is bifurcated into 
two distinct stages, each grounded in a combination of geometric and computational strategies.

In the initial stage, our methodology focuses on the broad localization of voids within the point cloud. This stage commences with the 
identification of principal points delineating the outermost envelope of the cloud, achieved through classical convex hull analysis. 
The vertices of the convex hull (CH) are then utilized as a basis for defining principal segments across the cloud.

Leveraging Delaunay Triangulation and Voronoi Diagrams in conjunction with the Minimal Distance Scoring (MDS) technique, which we present 
here, we systematically identify the set of segments that exhibit the highest likelihood of intersecting the void. This technique integrates 
spatial metrics derived from the triangulation and diagrams to score each segment. We then search for potential first-order Maximal Internal 
Envelope (MIE) points ($p_i^{MIE} \in P$), based on their proximity and angular relations to these segments.

The second stage adopts an iterative refinement process, building upon the foundational dataset of MIE points from the first stage. 
Each iteration ("order") enhances the structural definition of the void, iteratively converges towards a nuanced representation of 
the void's shape. 

This 'higher-order' approach, while computationally intensive, allows for a progressively more detailed and accurate delineation of the 
void's geometry. 
By iteratively leveraging the MIE points, the methodology converges towards a nuanced and precise representation of the void's fine shape 
characteristics, thereby encapsulating the complexity and subtlety inherent in the spatial structure of the point cloud.

\subsection{Void Identification}

The methodology for identifying voids within a two-dimensional point cloud involves a series of geometric and computational steps. 
We begin by constructing the set of segments connecting all paired Convex Hull vertices, applying the SDS algorithm to assign a Voidness 
Score parameter to each of the segments to find the segment of maximal likelihood to penetrate the void. We then apply the Delaunay 
Triangulation and Voronoi Diagrams analysis to identify the optimal point on that segment to represent the point who is considered to 
be the deepest inside the void.
The process involves the following key stages:

\subsubsection{Minimal Distance Scoring (MDS)}
\textbf{Definition:} The Minimal Distance Scoring (MDS) function evaluates the likelihood of a line segment being within a void. 
The MDS for a segment \(L_{ij}\) connecting points \(p_i\) and \(p_j\) in \(P\) is defined as:

\begin{equation}
    \text{MDS}(L_{ij}) = \frac{1}{|P \setminus P_{CH}|} \sum_{p_k \in P \setminus P_{CH}} d(L_{ij}, p_k),
	\label{MDS-V Scorring Eq.}
\end{equation}
where \(d(L_{ij}, p_k)\) is the distance from \(L_{ij}\) to a point \(p_k\) in \(P \setminus P_{CH}\).

The principle approach used here for identifying the void is as follows:

\begin{itemize}

    \item \textbf{Convex Hull Identification}: Using an algorithm such as Graham's Scan or Jarvis's March~\cite{Preparata1977ConvexHulls}, 
	the Convex Hull \(P^{CH}\) of \(P\) is identified. 

    \item \textbf{Starting Point Selection}: A point $P^{CH}_i$ is selected from the subset $P^{CH}$. This selection can be 
	randomized within $P^{CH}$ or based on a geometric heuristic, such as the point's centrality or distance to other points in $P^{CH}$.

    \item \textbf{Segment Creation and Analysis}: For each \(P^{CH}_i\), segments \(L_{ij}\) connecting it to other points in \(P^{CH}\) 
	are created. These segments are analyzed for their potential to intersect the internal void.
	
    \item \textbf{Voidness Score Calculation}: The voidness score \(V(L_{ij}) = \text{MDS}(L_{ij})\) (see Equation \ref{MDS-V Scorring Eq.}) 
	for each segment \(L_{ij}\) is calculated as the average distance to points in \(P \setminus P^{CH}\).
    The segment with the lowest voidness score is selected as \(L^{best}_{ij}\) and is considered to have a greater potential to be close to 
	or within the internal void.
	
\end{itemize}

\subsubsection{Delaunay-Voronoi (DV) point selection}

Using Delaunay Triangulation and Voronoi Diagrams, we select the optimal point on \(L^{best}_{ij}\) that best represents the area with 
the lowest density in the point cloud.

\textbf{Delaunay Triangulation:} Given a set of points $P$ in a two-dimensional Euclidean space, a Delaunay triangulation for $P$ is a 
triangulation DT($P$) such that no point in $P$ is inside the circumcircle of any triangle in DT($P$).

\vspace{0.4cm}

\textbf{Properties:}
\begin{itemize}
    \item Maximizes the minimum angle of all the angles of the triangles in the triangulation, avoiding narrow triangles.
    \item The triangulation is unique if no four points are cocircular.
    \item Each triangle's circumcircle contains no other points from $P$ in its interior.
\end{itemize}

\textbf{Voronoi Diagrams:} The Voronoi diagram for a set of points $P$ in a two-dimensional space partitions the plane into regions where each 
region corresponds to a point in $P$. For each point $p_i \in P$, its corresponding Voronoi cell, denoted as $V(p_i)$, consists of 
all points closer to $p_i$ than to any other point in $P$.

\vspace{0.4cm}

\textbf{Properties:}
\begin{itemize}
    \item Each Voronoi cell $V(p_i)$ is a convex polygon.
    \item The diagram is the dual graph to the Delaunay triangulation of $P$.
    \item The edges of the Voronoi diagram are perpendicular bisectors of the edges of the Delaunay triangulation.
\end{itemize}

The point that best represent the area with lowest density of cloud point is selected as the DV point.

\subsubsection{Schematic Algorithm}

\begin{figure}[H]
\begin{center}
\includegraphics[width=9cm]{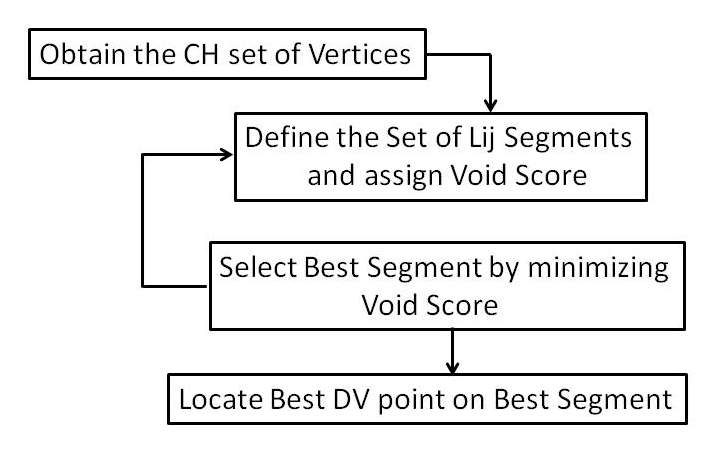}
\caption{Schematic Algorithm for Void Identification.}
\label{Algorithm_1 Figure}
\end{center}
\end{figure}

The process is schematically represented below \ref{Algorithm_1 Figure}, detailing the computational steps for void identification.

\begin{table}[H]
\centering
\caption{Algorithm for Void Identification}
\begin{tabular}{rl}
\textbf{Step} & \textbf{Action} \\
1: & Identify Convex Hull \(P^{CH}\) of \(P\) \\
2: & Select a starting point \(P^{CH}_i\) from \(P^{CH}\) \\
3: & For each \(P^{CH}_i\), create and analyze segments \(L_{ij}\) \\
4: & Calculate MDS(\(L_{ij}\)) for each segment \\
5: & Select \(L^{best}_{ij}\) with minimum MDS value \\
6: & Apply Delaunay Triangulation and Voronoi Diagrams to select DV point on \(L^{best}_{ij}\) \\
\end{tabular}
\label{table:algorithm 1}
\end{table}

\vspace{0.5cm}

\subsection{Void Shape Construction}

\begin{figure}[H]
\begin{center}
\includegraphics[width=9cm]{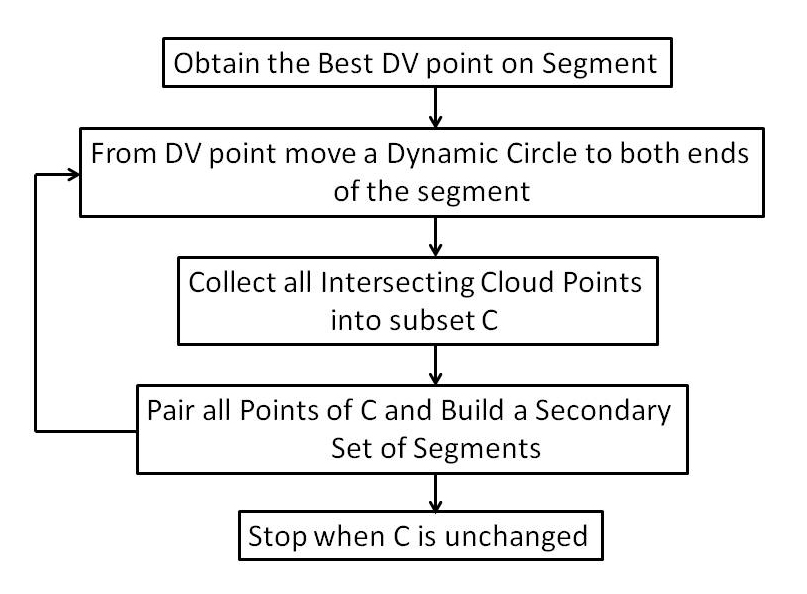}
\caption{Schematic Algorithm for Void Shape Construction.}
\label{Algorithm_2 Figure}
\end{center}
\end{figure}

The final stage involves constructing the shape of the void using an expanding polygon methodology, initiated from the 
segment \( L^{best}_{ij} \) identified by the MDS algorithm. This approach combines computational geometry techniques with 
iterative algorithms to accurately model the void's shape.
A schematic representation of the above is shown for clarity below \ref{Algorithm_2 Figure}.

\subsubsection{Polygon Construction}
\begin{itemize}

    \item \textbf{Vertex Selection:} The vertices of the expanding polygon, denoted by the set $C \subseteq P$, are chosen based on their 
	proximity to \( L^{best}_{ij} \). 
    The process begins with a minimal set of points in \( C \) and expands by adding new vertices from \( P \).	
	
    \item \textbf{Initial Configuration:} The process initiates with an imaginary circle centered at the DV point on the 
	segment \( L^{best}_{ij} \). This circle gradually expands in radius until it first intersects with points in the cloud. 
	The center of the circle then traverses along the segment \( L^{best}_{ij} \), extending towards both endpoints is predefined 
	step sizes (controlled by parameter $3 \leq k \leq n$). Throughout this traversal, points that are first intersecting with the 
	expanding circle are accumulated into the set \( C \), forming the initial vertex set for the polygon. This procedure is 
	systematically repeated for all segments \( L_{ij} \) associated with \( P^{CH}_i \), ensuring comprehensive coverage and 
	inclusion of relevant points in the vicinity of the void.

    \item \textbf{Growth Algorithm:} Starting with the initial set of Maximal Internal Envelope (MIE) points in \( C \), the algorithm 
	enters an iterative phase, where each iteration is marked by an increasing "Order" of complexity and refinement. 
	At each iteration, the algorithm constructs all possible segments connecting the vertices currently in \( C \). For each of these 
	segments, the imaginary circle process is reapplied. This process involves centering an imaginary circle on various points along 
	each segment and expanding it until it first intersects with a new point in the cloud. The newly intersected point, not previously 
	included in \( C \), is then added to the MIE collection. This iterative expansion and addition of points to \( C \) continues, 
	thereby progressively refining and elaborating the polygonal representation of the void’s boundary with each successive order.

    \item \textbf{Constraint by Points in \( P \):} As the polygon expands, its growth is influenced by the fixed points in \( P \). 
	These points act as constraints, guiding the polygon's expansion. When a potential expansion intersects with a point in \( P \), 
	that point becomes a candidate to be added as a new vertex to the polygon. This process ensures that the expanding polygon adapts 
	its shape to encompass the void, respecting the spatial distribution of \( P \).

    \item \textbf{Intersection Condition:}
       \begin{itemize}
           \item A point $p \in P$ is considered to intersect with the expanding polygon if the distance from $p$ to the nearest line segment of 
	       the polygon is less than a predefined constant $\delta$.
          \item Points satisfying this condition are incorporated into the set $C$, altering the polygon's shape.
       \end{itemize}

    \item \textbf{Adaptive Shape Modification:} The shape of the polygon at each stage of expansion is a collection of line segments 
	connecting adjacent vertices in \( C \). This means the polygon is always a closed, piecewise-linear path. The algorithm ensures 
	that with each expansion step, the polygon remains a valid geometric shape. It dynamically adjusts the polygon by adding new 
	vertices, removing redundant ones, or altering the connections between vertices to best represent the boundary of the void.

    \item \textbf{No Internal Points Assumption:} It is assumed that no point in $P$ reside insides the expanding polygon. Under this 
	assumption, a void cannot be defined as such if it contains points within.	

    \item \textbf{Computational Implementation:} Implementing these dynamics requires an iterative process where the polygon's shape is 
	continuously updated. Each iteration involves checking for potential expansions, evaluating constraints from \( P \), and 
	updating \( C \) and the polygon's edges accordingly. 
	Data structures, such as priority queues \cite{Kumar:2020}, might be used to efficiently manage potential expansion candidates based 
	on their geometrical locations relative to cloud points $p_i \in P$.

\end{itemize}

Through these dynamics, the expanding polygon evolves in a manner that is both controlled by an internal growth logic and responsive to 
the external environment defined by the points in \( P \). This dual influence ensures that the polygon effectively and accurately 
delineates the shape of the void.

\subsubsection{Refining Criterion}
\begin{itemize}
    \item Define a 'Refining' parameter $k = 1 / \delta$ of the expanding polygon, representing the number of steps along $L^{best}_{ij}$. 
	This parameter is employed by the algorithm to facilitate the pursuit of an additional intersecting point, thereby enhancing the 
	refinement process.
    \item The refining ranges from a minimum of 3 with a practical upper limit of $n$, the number of points in $P$.
    \item The expansion process is observed through increasing orders, providing a measure for the growth and adaptation of the expanding 
	polygon to the void’s shape.
\end{itemize}

\begin{table}[h]
\centering
\caption{Algorithm for Void Shape Construction}
\begin{tabular}{rl}
\textbf{Step} & \textbf{Action} \\
1: & Identify segments \(L^{best}_{ij}\) with minimum MDS value. \\
2: & Along each segment \(L^{best}_{ij}\), define a dynamic imaginary circle that expands from the DV point. \\
3: & Identify Maximal Internal Envelope (MIE) points from \(P\) that intersect with the expanding circle. \\
4: & Incorporate these intersecting points into the subset \(C\). \\
5: & Update the set \(C\) and repeat the process for the new segments formed. \\
6: & Terminate the iterative process when the addition of new points to \(C\) ceases. \\
\end{tabular}
\label{table:algorithm 2}
\end{table}

\vspace{0.5cm}

\subsection{Complexity Analysis}

Given a 2D point cloud with $n$ points, we aim to analyze the complexity of the process as described above for void identification
and shape construction:

\begin{enumerate}
    \item Compute the convex hull of the point cloud.
    \item Generate all unique pairs of vertices on the convex hull.
    \item Generate pairs from a subset of points (approximately 1/10th the size of the original set) and repeat this step up to a 
	maximal order of 3.
\end{enumerate}

\paragraph{Detailed Considerations}

\begin{itemize}

    \item \textbf{Convex Hull Calculation:} The convex hull of a point cloud is computed, which typically has a complexity of $O(n \log n)$.

    \item \textbf{Pair Generation of Convex Hull Vertices:} Assuming the convex hull has approximately $\frac{n}{10}$ vertices, the number 
	of pairs generated is $\binom{\frac{n}{10}}{2} \approx \frac{n^2}{200}$. Hence, the complexity for this step 
	is $O\left(\frac{n^2}{100}\right)$.

    \item \textbf{Subset Pair Generation:} For a subset of points similar in size to the convex hull vertices, the complexity for pair 
	generation remains $O\left(\frac{n^2}{100}\right)$. However, for moderate-sized $n$, this quadratic term is manageable and does not 
    drastically impact performance. 

    \item \textbf{Repetitions:} Repeating the pair generation for subsets twice more adds a cumulative complexity 
	of $2 \times O\left(\frac{n^2}{100}\right)$.

\end{itemize}

\paragraph{Overall Complexity}
Combining all steps, the total complexity is:

\begin{equation}
    O(n \log n) + 4 \times O\left(\frac{n^2}{100}\right)
\end{equation}

This expression shows that while the convex hull computation is $O(n \log n)$, the pair generation steps, particularly for larger $n$, 
contribute a significant quadratic component to the overall complexity.
Given the moderate size of $n$, this complexity is generally well-handled by contemporary computational resources.

The study underscores the need for optimizing such algorithms, particularly for applications dealing with extensive spatial datasets.
In scenarios involving moderate-sized 2D point clouds however, the complexity of the algorithms used for convex hull computation and pair 
generation, does not pose significant challenges. This above analysis demonstrates that such tasks are computationally feasible within the 
constraints of typical processing capabilities. 

\section{Results - Void Identification and Shape Construction}
\begin{figure}[H]
\begin{center}
\includegraphics[width=7cm]{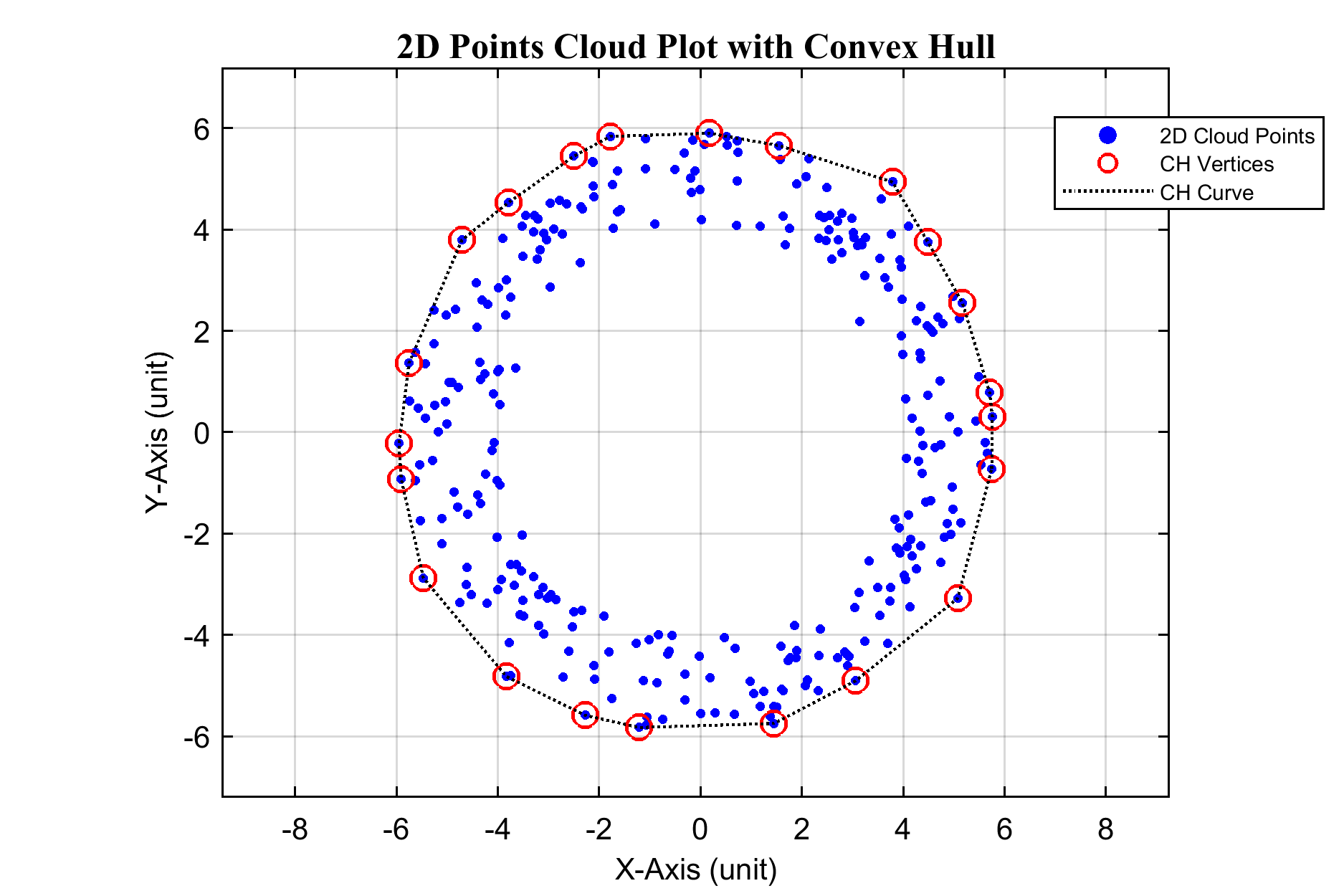}
\includegraphics[width=7cm]{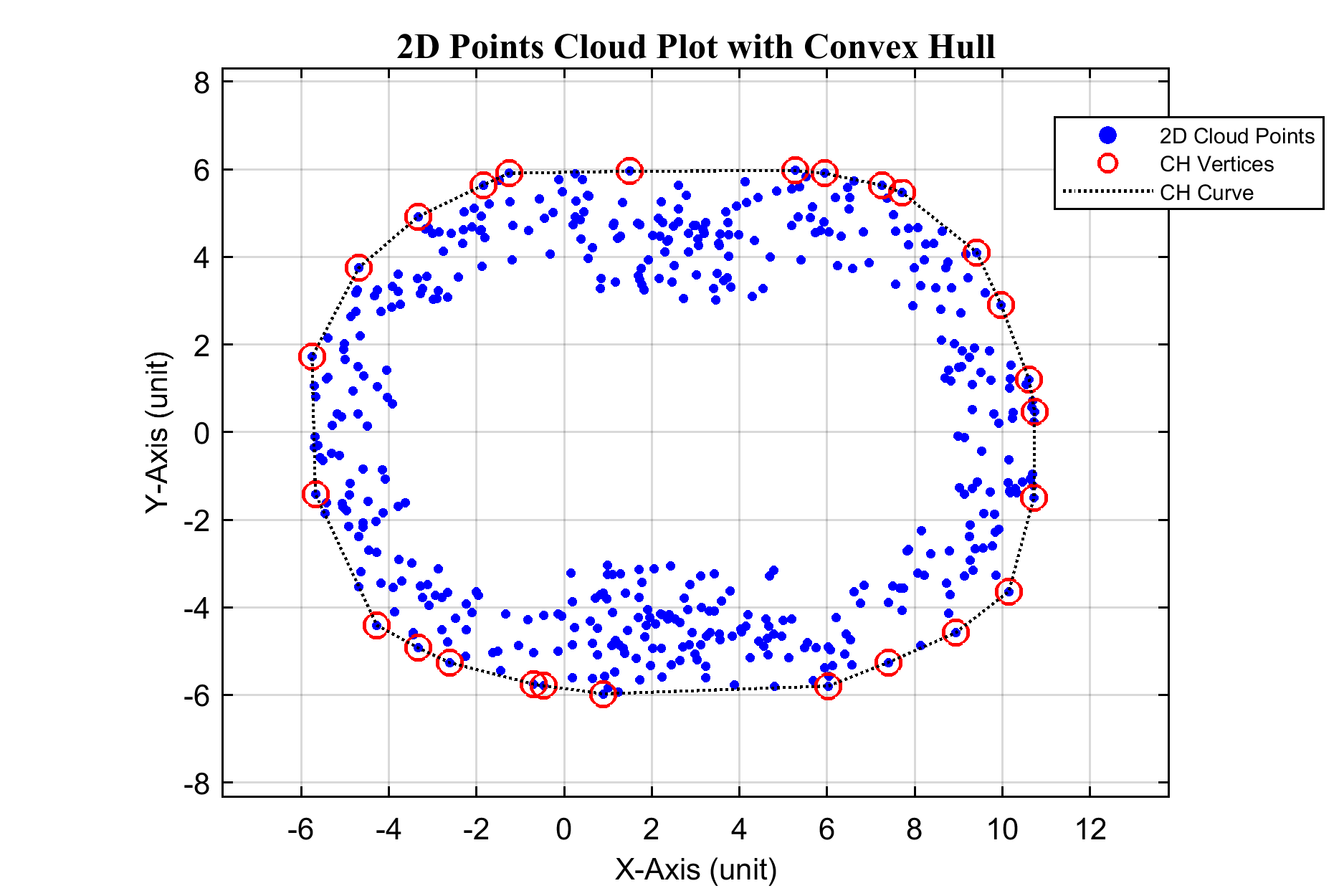}
\caption{The 2D point distributions of a shell-like circle (left) and a moderately detailed shape (right).}
\label{Cloud and CH figure}
\end{center}
\end{figure}

In the present study, two distinct ensembles of 2D point sets were employed to illustrate the construction of the Maximal Internal 
Envelope (MIE) shape: a canonical circle with random distribution along the circumference, and a pair of congruent circles offset by 
half a radius with an internal void. The Convex Hull vertices for each case were computed, as shown in Figure \ref{Cloud and CH figure}.

\begin{figure}[H]
\begin{center}
\includegraphics[width=7cm]{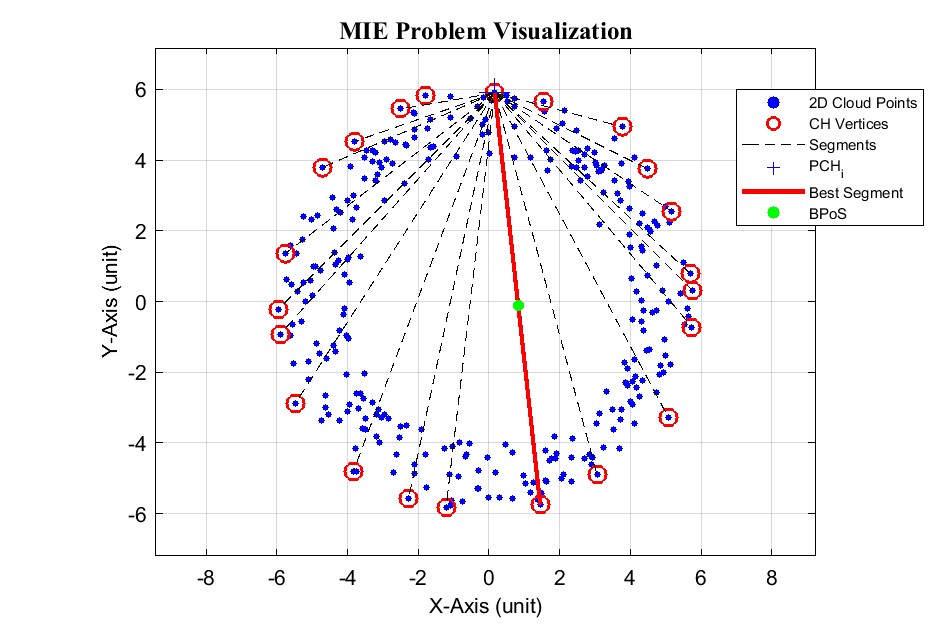}
\includegraphics[width=7cm]{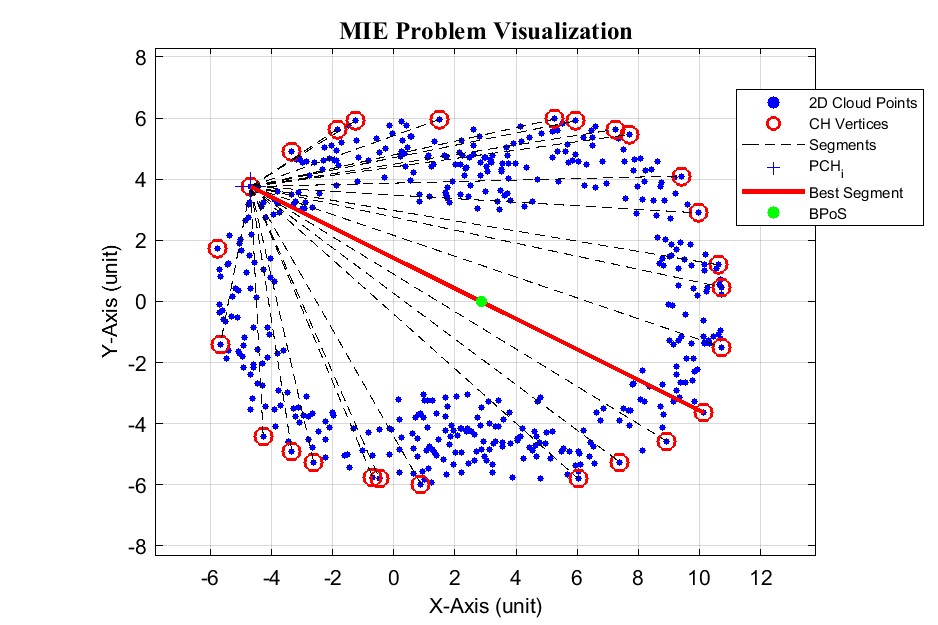}
\caption{Example of best segments selection based on MDS for the two clouds. Every segment is connecting CH pairs of the respective 2D 
points distribution.}
\label{Best segments selection figure}
\end{center}
\end{figure}

\begin{figure}[H]
\begin{center}
\includegraphics[width=7cm]{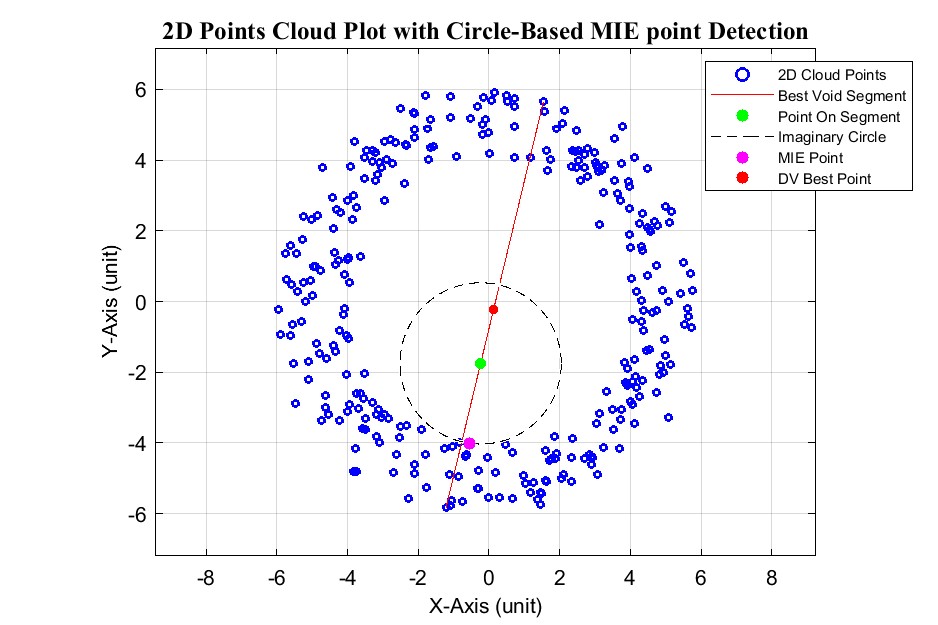}
\includegraphics[width=7cm]{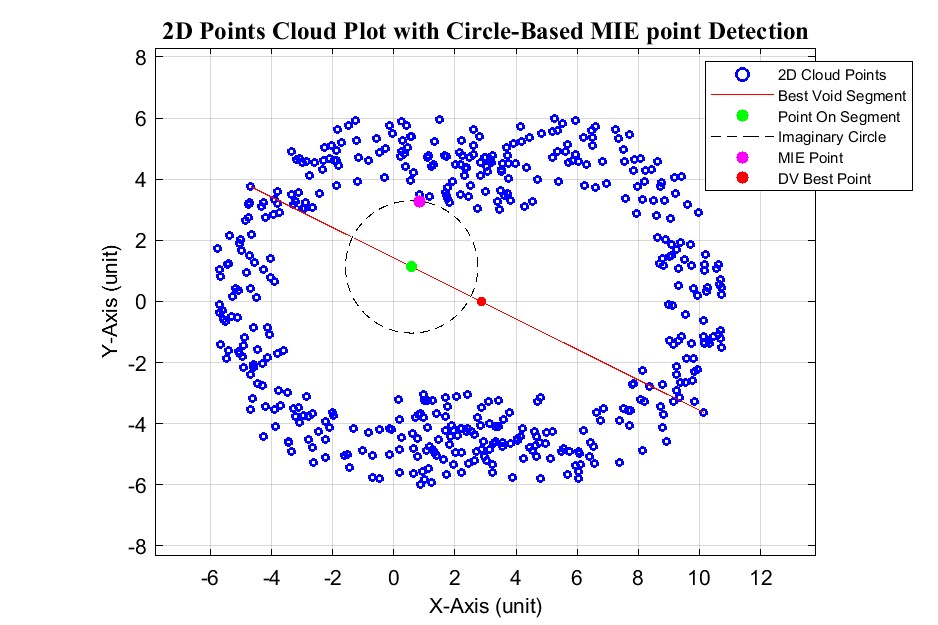}
\caption{Example of MIE points identification based on best segment and VD starting point on segment for the interion void polygon construction.}
\label{Imaginary Circle figure}
\end{center}
\end{figure}

We initiated by selecting optimal segments from the Convex Hull points, aiming to identify those traversing the void. Delaunay triangulation 
and Voronoi diagrams were then used to determine the starting point for the envelope construction (see in 
Figure \ref{Best segments selection figure}). 
The process involved projecting an imaginary circle from the Delaunay-Voronoi (DV) point along the selected segment. Points 
intersecting with this circle, as shown in Figure \ref{Imaginary Circle figure}, were included in the subset \( C \subseteq P \). 
These points, representing the Maximal Internal Envelope (MIE), were instrumental in outlining the curve that encapsulates the 
internal void, thereby providing a detailed and quantifiable representation of its spatial characteristics.

\begin{figure}[H]
\begin{center}
\includegraphics[width=7cm]{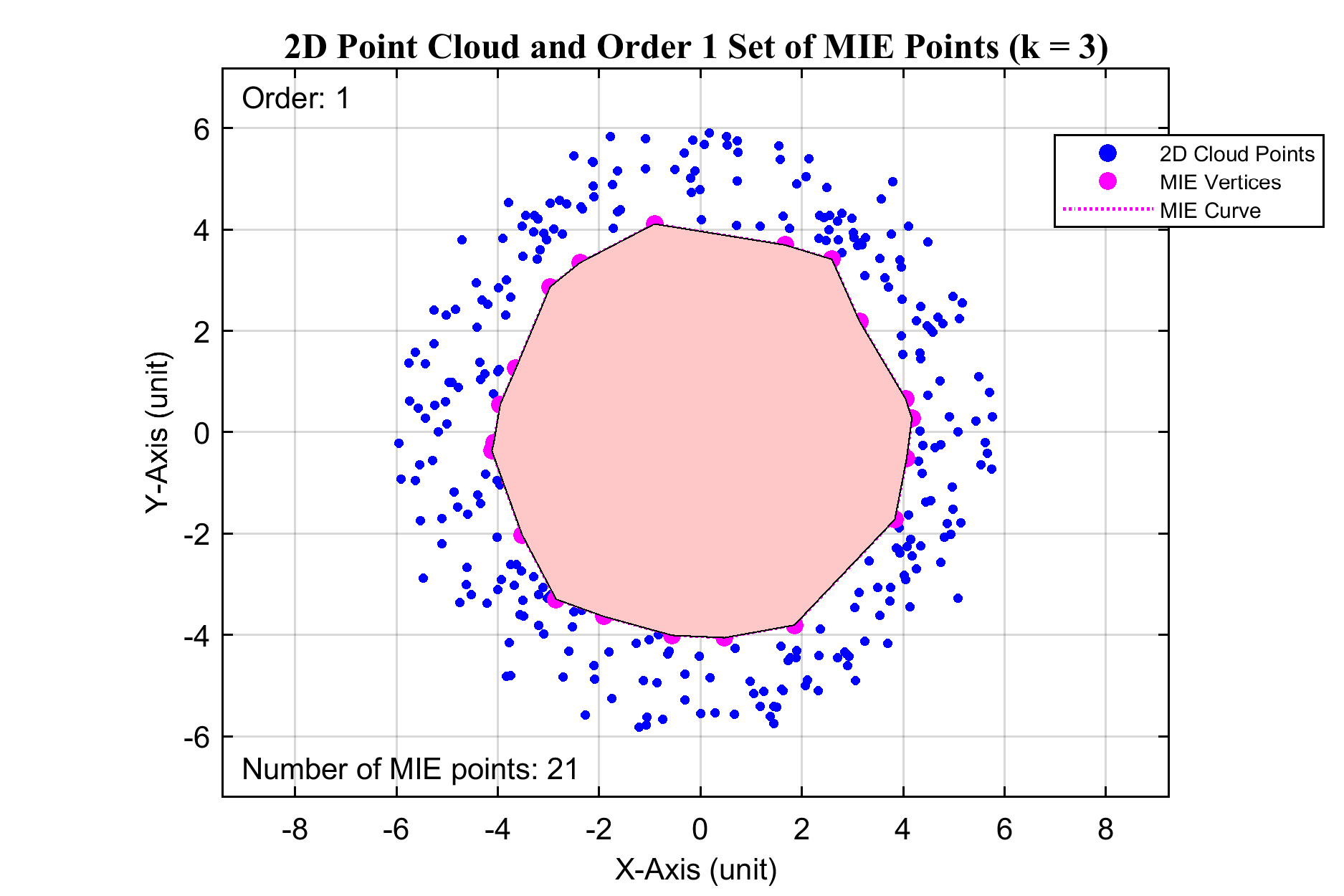}
\includegraphics[width=7cm]{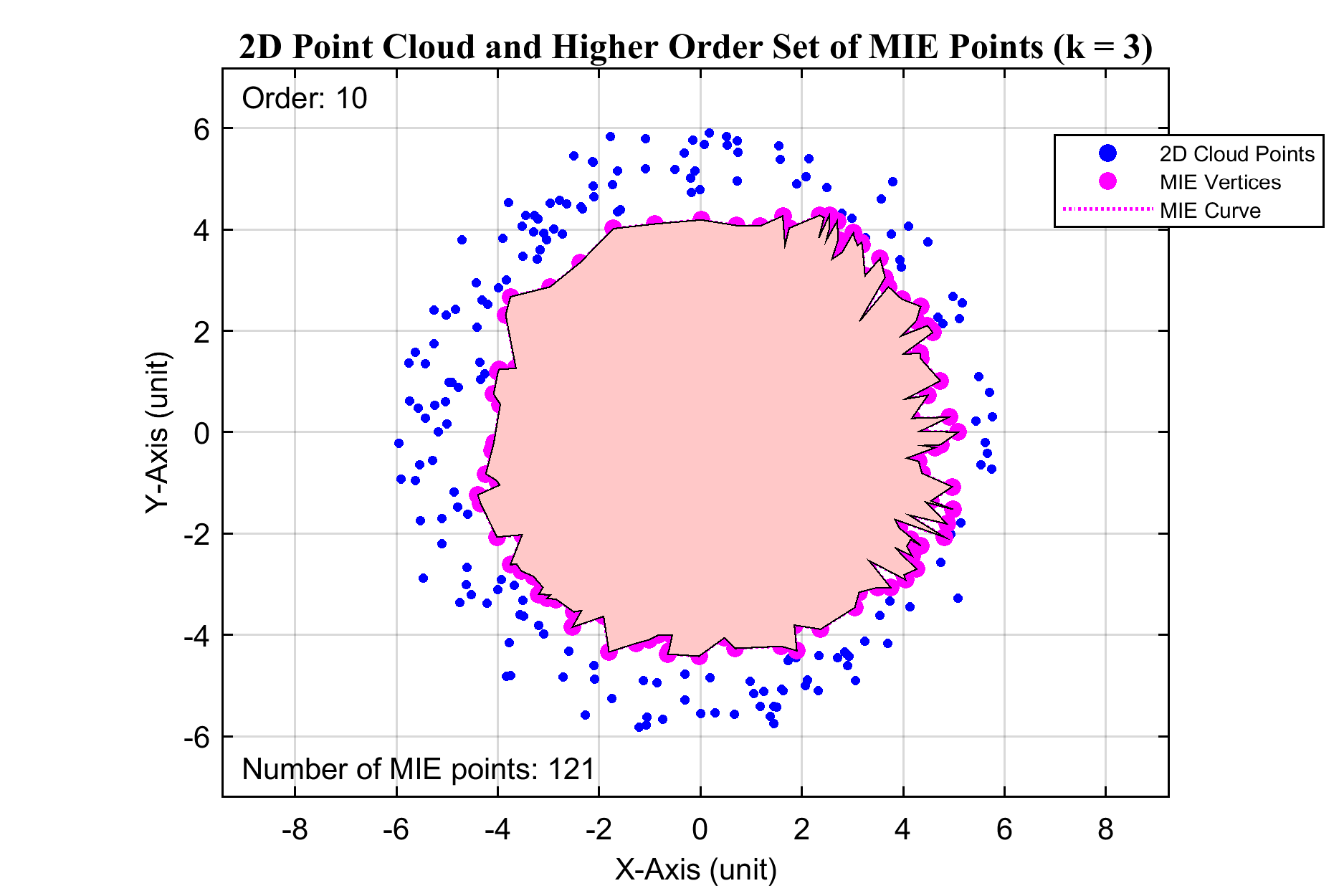} \\
\includegraphics[width=7cm]{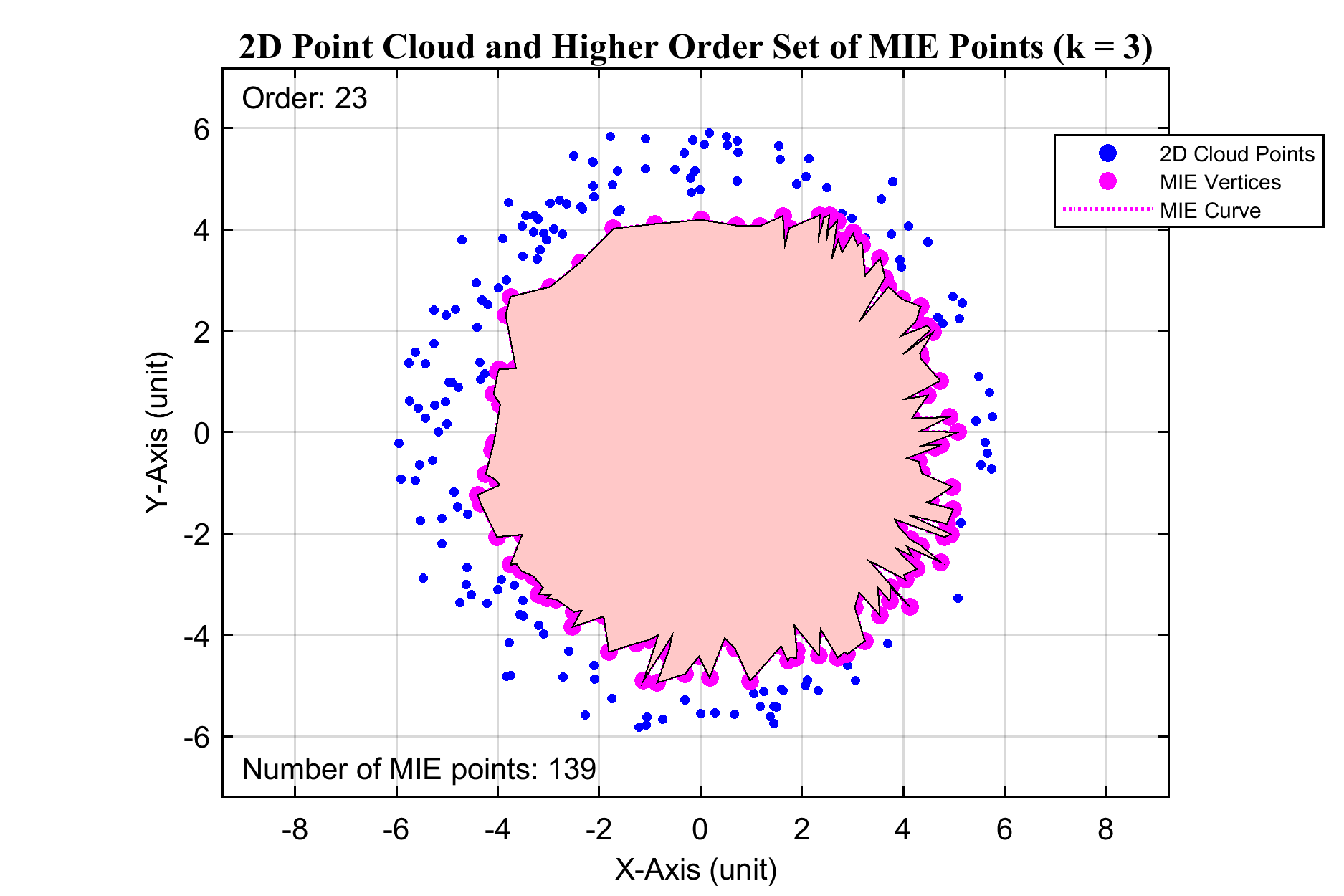}
\includegraphics[width=7cm]{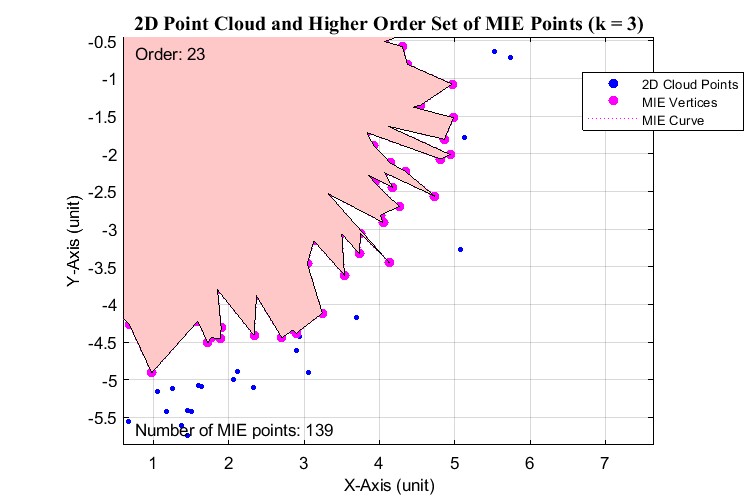} 
\caption{The final maximal interior void in the single-circle shape with different orders for $k=3$.}
\label{k is 3 Orders 1 10 23 z23  C1_Figure}
\end{center}
\end{figure}

\begin{figure}[H]
\begin{center}
\includegraphics[width=7cm]{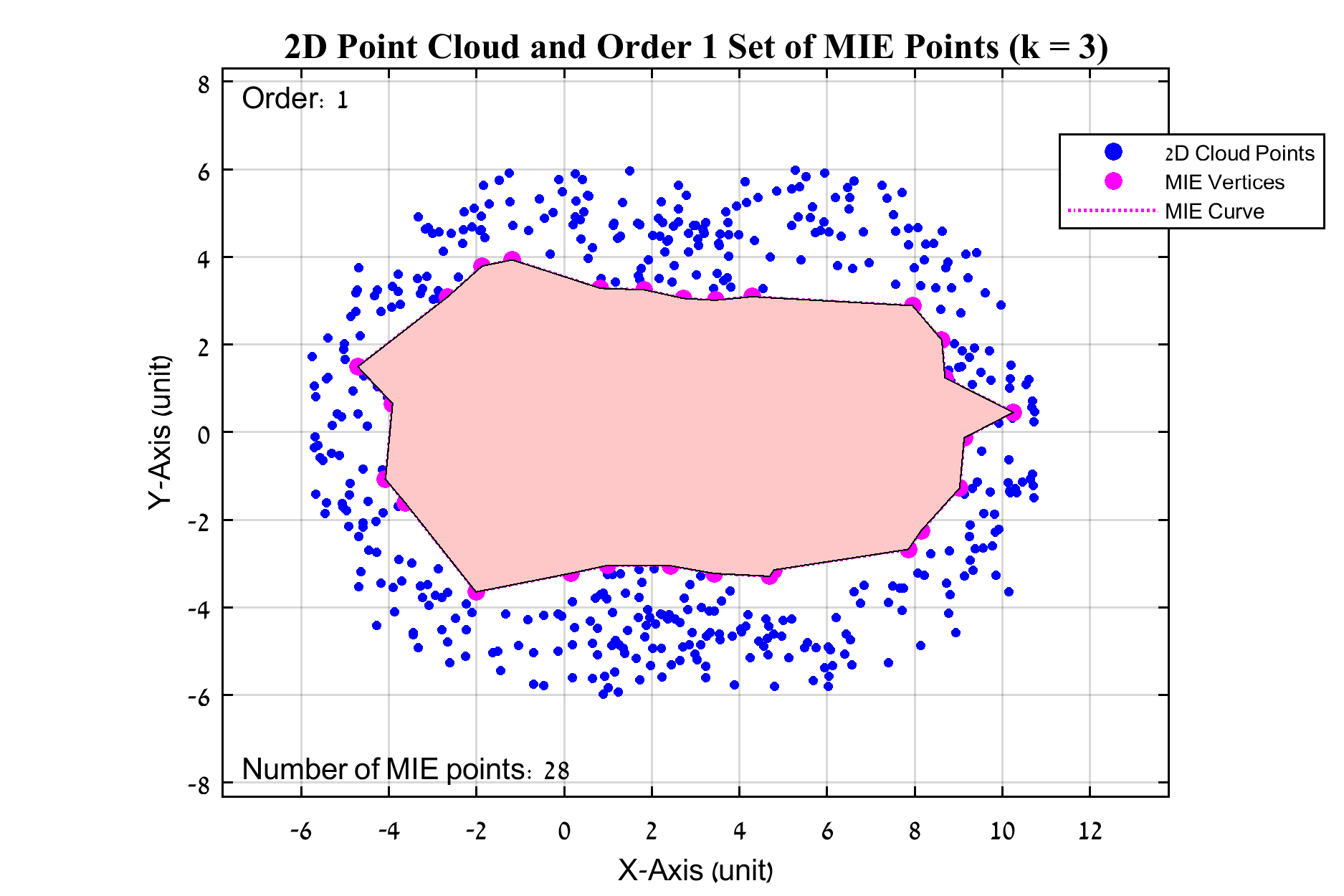}
\includegraphics[width=7cm]{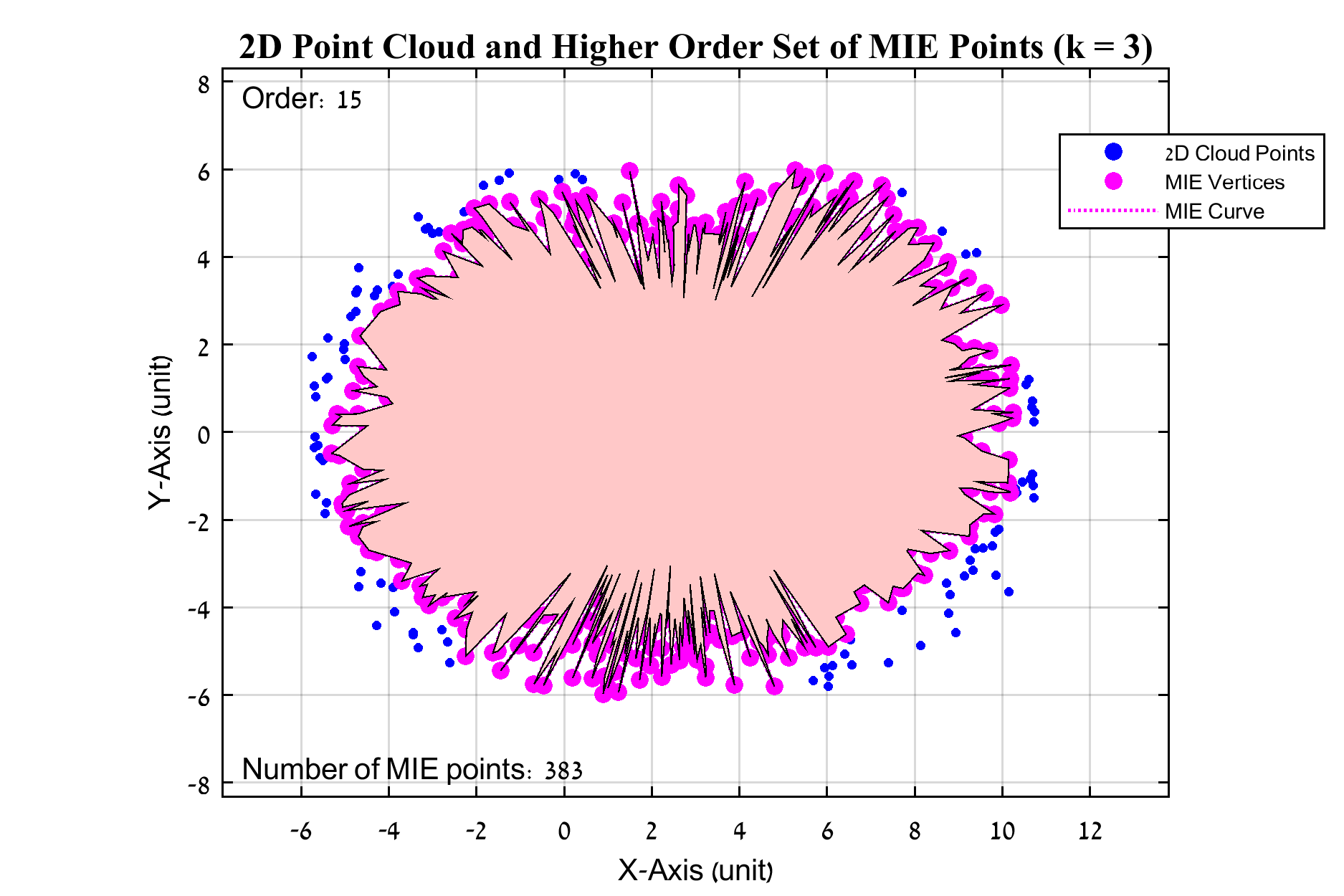} \\
\includegraphics[width=7cm]{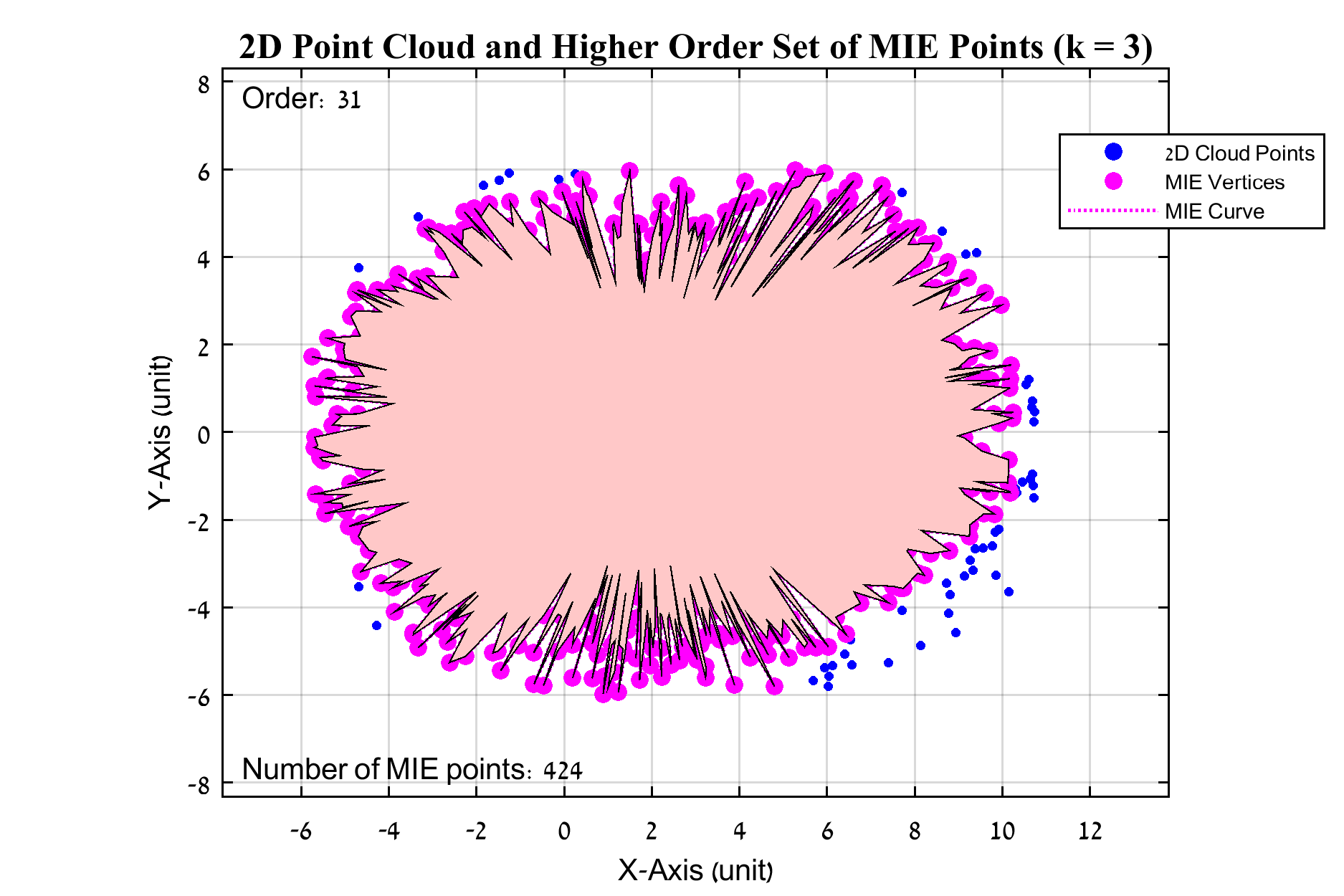}
\includegraphics[width=7cm]{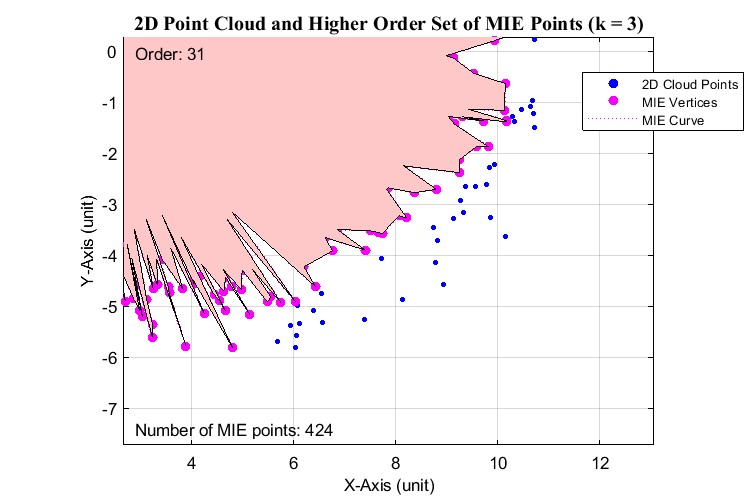} 
\caption{The final maximal interior void in the dual-circles shape with different orders for $k=3$.}
\label{k is 3 Orders 1 15 31 z31  C2_Figure}
\end{center}
\end{figure}

The resultant shapes of the Maximum Internal Envelope (MIE) for the two aforementioned scenarios are presented in a comparative analysis, 
catering to distinct settings of the parameter $k$ and increasing orders of calculations. Specifically, we exhibit the MIE configurations 
for $k=3$ and $k=31$, \footnote{These values were selected arbitrarily to showcase the approach's versatility and scalability under 
various conditions.} and different calculation orders. Table \ref{tab:data-summary} shows the convergence orders for the cases 
simulated here.
The respective visual representations of these MIE shapes are illustrated in Figures \ref{k is 3 Orders 1 10 23 z23  C1_Figure} 
for the circle-shaped cloud, and \ref{k is 3 Orders 1 15 31 z31  C2_Figure}  for the dual-circles-shaped cloud with $k=3$.

\begin{table}[h]
\centering
\caption{Data Summary}
\label{tab:data-summary}
\begin{tabular}{lcc}
\hline
\textbf{Shape} & \textbf{k-value} & \textbf{Orders} \\ \hline
1 Circle       & 3                & 23              \\
1 Circle       & 31               & 18              \\
2 Circles      & 3                & 31              \\
2 Circles      & 31               & 22              \\ \hline
\end{tabular}
\end{table}

\begin{figure}[H]
\begin{center}
\includegraphics[width=7cm]{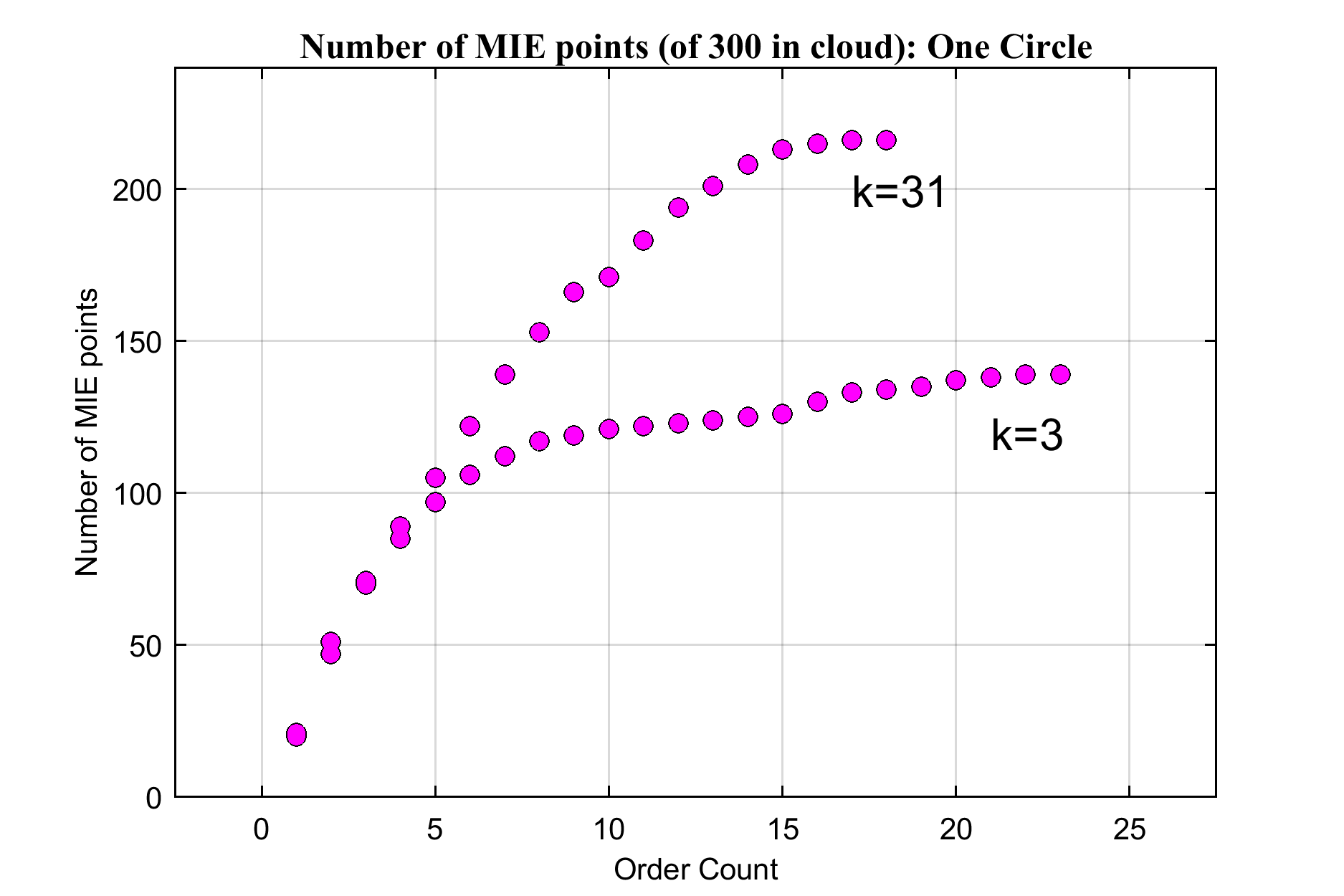}
\includegraphics[width=7cm]{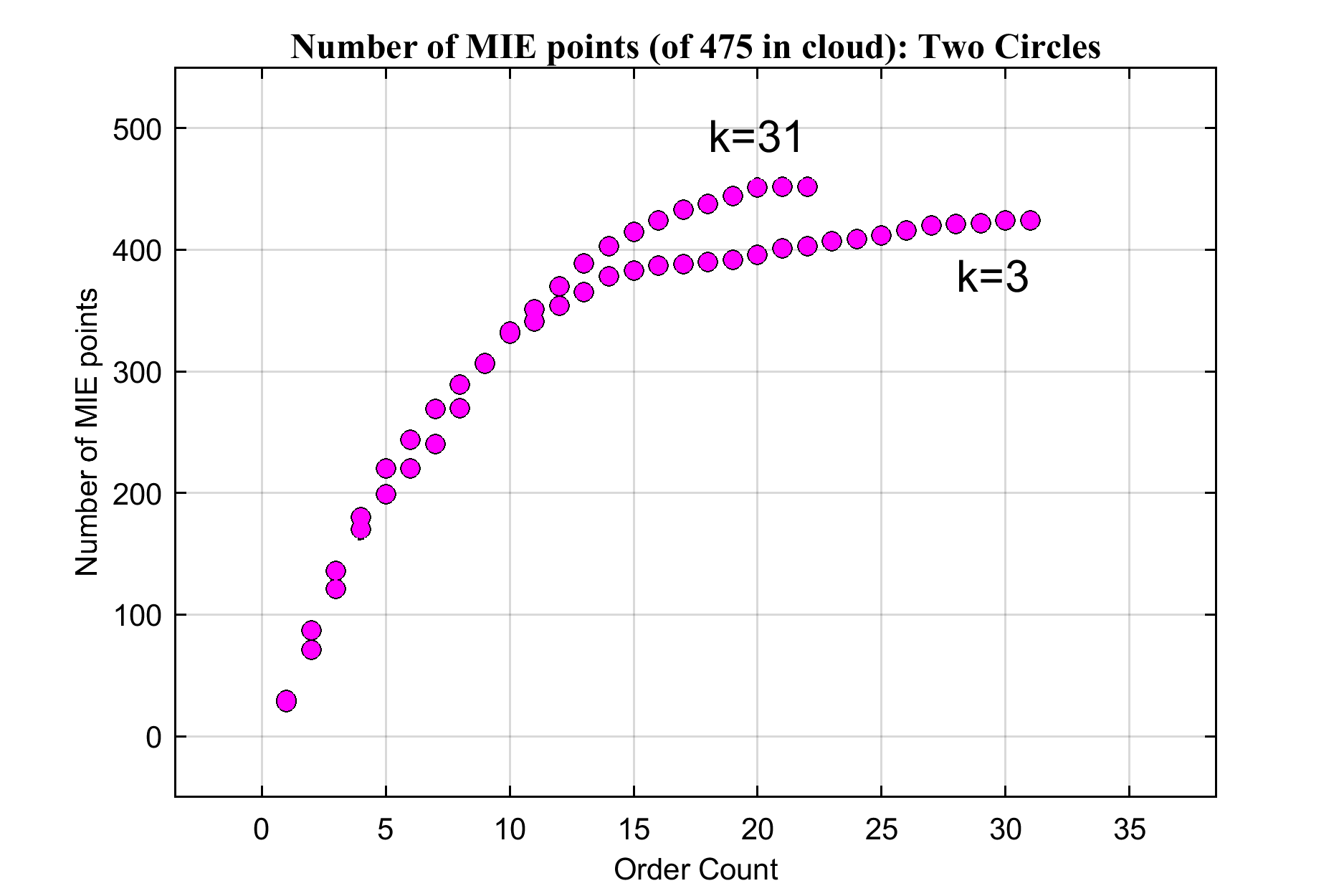}
\caption{Number of MIE points as a function of the calculation order for the dual-circles shape.}
\label{MIE_k C12_Figure}
\end{center}
\end{figure}

The figures in \ref{MIE_k C12_Figure} display the number of MIE points identified at various orders for for $k=3$ and $k=31$
in both the single-circle and dual-circle scenarios. These values are derived from the full spectrum of orders (iterations) 
conducted during our simulations.

In the course of our study, several key observations emerged that merit detailed consideration:

\begin{itemize}
    \item \textbf{Precision in Internal Void Identification:}
    The methodology employed in our study has demonstrated high accuracy in the delineation of internal voids. These voids were not only 
	clearly identified but also faithfully represented in terms of their visual and geometric properties. 
	This fidelity is crucial, as it underscores the reliability of our approach in capturing the intrinsic features of voids within 
	different geometrical structures.

    \item \textbf{Efficacy of Convex Hull Points in Segment Definition:}
    The utilization of convex hull points as the basis for defining segments has proven to be effective. 
	Nonetheless, it is pertinent to acknowledge that exploring a broader spectrum of segment definition methodologies could 
	potentially enhance the granularity and intricacy of the void structure revelation. This exploration could lead to a more 
	nuanced understanding of the internal geometry of complex shapes.

    \item \textbf{Impact of Parameter \( k \) on Void Structure Refinement:}
    Our findings clearly indicate that increasing the value of \( k \) substantially improves the refinement of the internal void 
	structure. 
	This enhancement is attributed to the increased number of steps taken along the segment with the highest probability of being 
	situated within the void. 
	The correlation between \( k \) and the detail in void representation is pivotal, as it demonstrates the tunability and adaptability 
	of our approach to different granularity levels.
	
    \item \textbf{Fine Coverage in High Order Calculations:}
    As expected, higher orders result in the identification of more MIE points. Importantly, these points consistently provide valid 
	coverage of the void's area in all cases, ensuring continuous spatial representation. The determination of the void's true shape, 
	however, is semantic and highly dependent on both the setting of \(k\) and the degree of the calculation order.
	
\end{itemize}

\section{Conclusions}

This study introduces a method for identifying and characterizing voids within two-dimensional point clouds, leveraging the synergistic 
integration of Delaunay triangulation, Voronoi diagrams, and Minimal Distance Scoring. The study highlights the method's efficacy in 
accurately delineating internal voids and its adaptability to diverse geometric configurations.

\subsection{Key Findings and Methodological Developments}
\begin{itemize}
\item The proposed method has demonstrated high accuracy in delineating internal voids, capturing their intrinsic geometric properties 
with hihg fidelity.
\item The effectiveness of using convex hull points for segment definition was affirmed, with the potential for further enhancement 
through exploration of alternative segment definition methodologies.
\item A significant correlation was observed between the parameter $k$ and the level of detail in void representation, illustrating 
the method's adaptability in refining the internal void structure.
\item The method exhibited robustness in reconstructing void shapes across different point cloud configurations, from simple symmetric 
circles to more complex structures.
\end{itemize}

\subsection{Limitations and Future Directions}
\begin{itemize}
\item The current study did not prioritize computational efficiency, presenting an avenue for future optimization, especially for handling 
large-scale datasets.
\item The method's efficacy in dealing with highly detailed voids remains to be tested and will form a critical part of our ongoing research.
\item Future work may also explore the integration of machine learning techniques to further refine the void identification process and 
automate certain aspects of the methodology.
\item The principles and algorithms utilized in this study suggest potential for generalization to higher-dimensional spaces. 
This aspect may enable broader applicability in analyzing more complex spatial data, warranting further investigation.
\item The computational complexity analysis reveals that while the convex hull computation is efficient, the quadratic nature of pair 
generation steps, especially for larger datasets, necessitates optimization for applications with extensive spatial data.
\end{itemize}

In conclusion, acknowledging the need for further research, this study concentrates on the broader field of geometric data interpretation 
and may offer a basis for future developments in various domains of spatial data processing for voids identification within spatial 
points distributions.

\section*{Declarations}
All data-related information and coding scripts discussed in the results section are available from the 
corresponding author upon request.

\renewcommand{\bibname}{References}

\end{document}